# Dynamic pilot wave bound states

Troy Shinbrot
Rutgers University
Piscataway, NJ 08854 USA
shinbrot@rutgers.edu

Abstract:

Motivated by recent parallels between classical bouncing droplet experiments and quantum bound states, we explore lessons that droplet experiments might teach us about the dynamics of quantum solutions. Since the classical experiment is periodically driven, we examine periodic driving of the integer spin Klein-Gordon equation. We find that an exact solution can be obtained, and surprisingly this solution necessarily produces *half-integer* orbital angular momentum. We stress that these findings are strictly mathematical; nevertheless this and other physical implications are intriguing, and suggest further study.

**Lead paragraph:**

**Bouncing oil droplets reproduce many features once thought to be unique to quantum mechanics, including interference effects and the emergence of discrete bound states. In this paper, we propose that the converse is also true: by applying periodic driving similar to that seen in classical bouncing droplet experiments, we can probe the dynamics of a quantum system of interacting particles. We find that exact solutions are possible, and these solutions provide potential insights into constraints on dynamical quantum states, including the surprising emergence of spin ½ states composed of spin-less particles.**

## 1: Introduction

Experimental studies of bouncing silicone oil droplets have demonstrated that fundamental features of quantum mechanics – including wave-like scattering and discrete bound states – can be produced in a classical setting[1,2]. These features involve behaviors that can be explained in part by the pilot-wave theory of de Broglie[3], in which the phase of a field thought to emanate from a particle affects the outcomes of quantum experiments. The significance of the intermediary field is visually apparent in bound states of bouncing droplets, as in the snapshot of a 3 droplet bound state shown in Fig. 1(a). In that panel, we see that the bouncing particles and their co-existing waves are dynamically synchronized: the binding force holding the particles in place results from waves produced by the particles themselves, and the bound state itself depends on the phase of the wave at the time when the particle makes contact with the surface[4].

Quantum bound states, on the other hand, intrinsically consider particles to absorb and emit intermediate field particles in quantized packets, but entirely ignore dynamical phase information. For example in the most elementary bound state studied in quantum mechanics – the Bohr model of the hydrogen atom – we are taught to define a Hamiltonian, $H = \frac{p^2}{2m} - \frac{e^2}{r}$, that contains no dynamical phase information. So an electron at any position and time always feels exactly the same attraction to the nucleus, as if an electron interacting with its intermediary photons generates forces that are completely static and continuous.

In this paper, motivated by droplet bound states where dynamical synchronization is clear-cut, we propose a dynamical Hamiltonian for a simple quantum bound state, where synchronization has to this point been routinely neglected. We show that the quantum formulation provides unexpected mathematical results. Among these is that zero spin particles can produce spin ½ bound states: a mathematical finding that could potentially have interesting physical consequences.

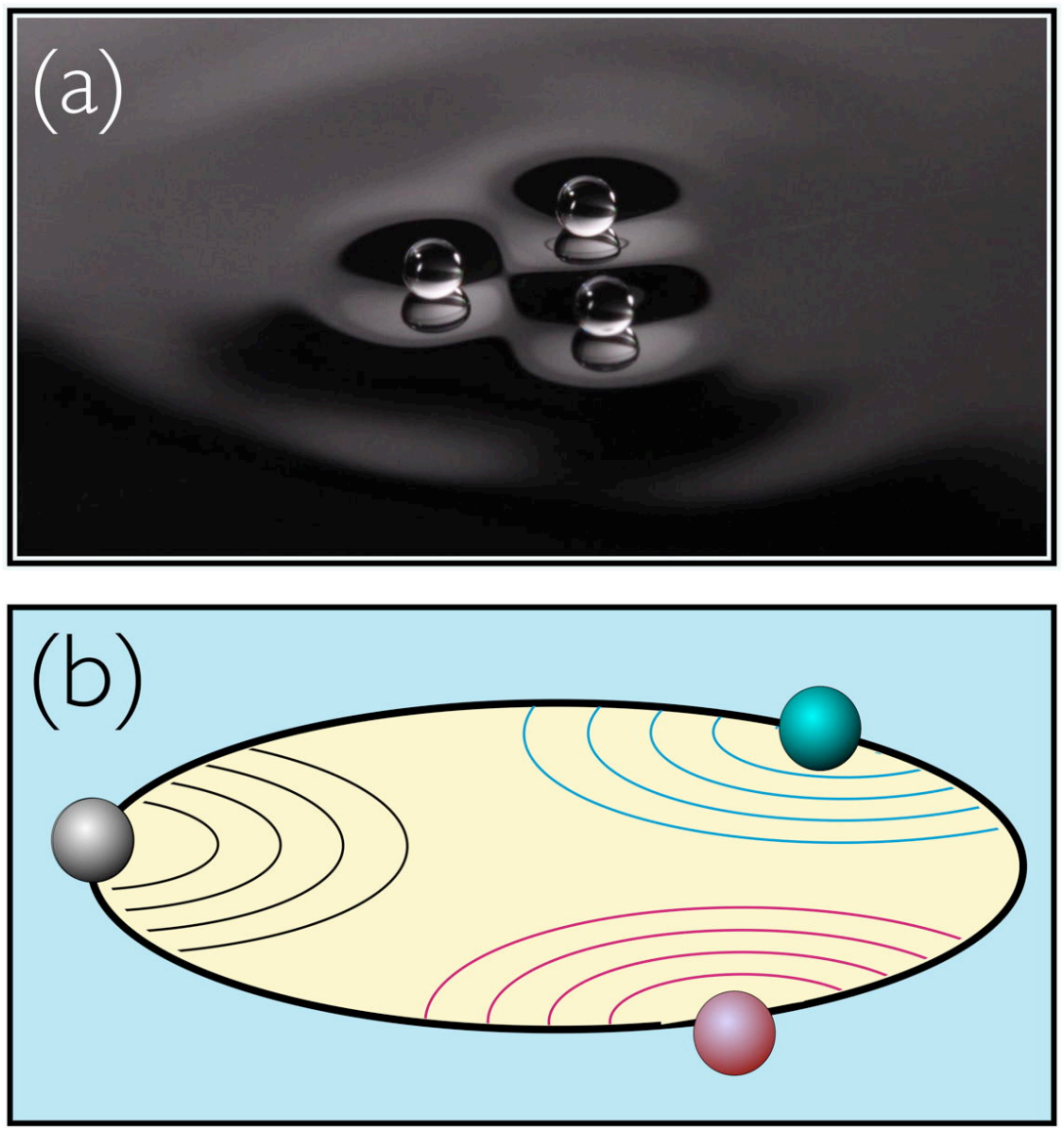

*Figure 1 – (a) Bound state of 3 oil droplets, image courtesy of Daniel Harris & John Bush, MIT. (b) Depiction of bound state of 3 particles, illustrating pilot waves emitted by each particle and absorbed by its neighbors.*

The quantum bound state that we choose to examine is perhaps the simplest mathematically imaginable, and is intended to resemble the bouncing droplet state of Fig. 1(a). In Fig. 1(b), we sketch a depiction of such a bound state, in which quantum particles periodically emit and absorb pilot waves so as to produce a harmonic potential:

$$H_I(r,\vartheta,t) = P(r)e^{i(k\vartheta - wt)}, \tag{1}$$

where $\vartheta$ and $t$ are angle and time variables, $k$ and $\omega$ are real constants, and $P(r)$ is a radial interaction, which is to be determined. Here the dynamical phase of the problem is made explicit by the term $(k\vartheta - wt)$. We will see that straightforward analysis puts constraints on the potential, and although $P(r)$ and $k$ are expressed in general terms, neither is free. We choose this form for the potential due to its analytic simplicity and because its explicitly complex form implies that a particle loses and gains energy periodically. This is in keeping both with the parallel to bouncing droplets – which lose and gain energy periodically – and with the notion that a bound state maintained by quantum intermediaries should do so as well.

To approach the problem, we will break it into two parts. In the first part, we will make use of the observation from experiments that bouncing droplets do become bound by a periodic driving field. We will neglect the question of how that field comes about in the quantum problem, and treat the periodic potential (1) as a given: a complete description of the influence of pilot wave dynamics acting on a quantum particle. This is not accurate to begin with because it neglects nonlinear interactions, but we will show that this simplification permits us to place constraints on exact solutions. We will also show, however, that the problem is overconstrained and so another degree of freedom is needed to complete the description. For the second part, we will incorporate such an additional degree of freedom by recognizing that a full solution requires feedback between pilot waves and particles. Short of producing a full nonlinear solution, we will construct an approximate form for the feedback by allowing energy to periodically be emitted and absorbed by a particle in an *ad hoc* way. We will show that this does admit an exact solution, and we will discuss some of its properties.

## 2: Klein-Gordon solution for a particle in a periodic potential

We seek to describe a quantum analog of a bouncing droplet in the periodic potential (1). A straightforward choice of model for this problem is the Klein-Gordon equation: this equation is conceptually simple and is generally applicable to symmetric (i.e. zero spin) particle dynamics. We will show that the resulting wavefunction can only exist with fixed *half*-integer orbital angular momentum. The fact that half-integer angular momentum can be generated from integer spin particles merely by including simple dynamics is remarkable; we emphasize here and subsequently that this and other results arise from purely mathematical considerations.

The Klein-Gordon equation in 2D is as follows[5]:

$$(p^2 + \mu^2 c^2)\varphi = \frac{1}{c^2}\left(i\hbar\frac{\partial}{\partial t} - H_I\right)^2 \varphi,$$

where $\mu$ is as usual mass and $p = -i\hbar\nabla$ is the momentum operator, so:

$$\nabla^2\varphi - \frac{\mu^2 c^2}{\hbar^2}\varphi = -\frac{1}{c^2}\frac{\partial^2}{\partial t^2}\varphi + \frac{2i}{\hbar c^2}H_I\frac{\partial}{\partial t}\varphi + \frac{i}{\hbar c^2}\,\varphi\frac{\partial}{\partial t}H_I - \left(\frac{H_I}{\hbar c}\right)^2\varphi\ . \tag{2}$$

When we insert the potential from Eq. (1), we find that the resulting equation is not separable by normal techniques. We can nonetheless proceed by attempting a solution of a doubly exponentiated, form:

$$\varphi = M(r)\ \exp\left\{i(\hat{k}\vartheta - \widehat{\omega}t) + f\, r^g \mathrm{e}^{i(k\vartheta - wt)}\right\}, \tag{3}$$

where $k, w, \hat{\kappa}, \hat{\omega}, f$ and $g$ are constants, and $M$ is a (second) unknown function of radius. We will discuss a different solution approach in the second part of this paper, but for the time being we point out that the role of the second exponential in Eq. (3) is to cause differentiation operations to bring down terms that can cancel with the exponential in the potential. Explicitly, we expand Eq. (2) and collect terms in like powers of $\exp[i(k\vartheta - wt)]$:

$$\begin{aligned} 0 = &\left\{M''(r) + \frac{1}{r}M'(r) - \left[\frac{w^2}{c^2} + \frac{\mu^2 c^2}{\hbar^2} + \frac{\hat{\kappa}^2}{r^2}\right] M(r)\right\} \\ &+ \left\{2gM'(r)r^{g-1} + \left[(g^2 - k^2 - 2fk\hat{\kappa})r^{g-2} - \frac{3w^2}{c^2}r^g - \frac{3wP(r)}{f\hbar c^2}\right] M(r)\right\} f \exp[i(k\vartheta - wt)] \\ &+ \left\{(g^2 - k^2)r^{2g-2} - \frac{w^2}{c^2}r^{2g} - \frac{2w}{f\hbar c^2}r^g P(r) + \frac{1}{f^2\hbar^2c^2}P^2(r)\right\} f^2 M(r) \exp[2i(k\vartheta - wt)], \end{aligned} \tag{4}$$

where we have set $\hat{\omega} = w$. This is not necessary, but it simplifies subsequent algebra, and has the physical interpretation of setting the frequency of the rotating wave, $\hat{\omega}$, equal to the frequency of the rotating potential, $w$. Each of the quantities in curly brackets in Eq. (4) must vanish independently, so we have transformed the PDE (2) into three simultaneous ODE's. A solution can now exist if all three ODE's can be made consistent.

Up to this point, the radial potential, $P(r)$, has been arbitrary. We can deduce from the last curly bracket in Eq. (4) that only one form of $P(r)$ will permit a solution, because this bracket contains several powers of $r$, each of which must cancel to permit a solution at all $r$. In particular, the bracket contains a single term in $r^{2g}$, which can only vanish if $P(r)$ contains a term in $r^g$. Likewise, the last bracket contains a term in $r^{2g-2}$, which can be cancelled if $P(r)$ contains a term in $r^{g-1}$ as well, so we try:

$$P(r) = P_g r^g + P_{g-1} r^{g-1}, \tag{5}$$

where $P_g$, and $P_{g-1}$ are undetermined coefficients. Expanding the last bracket in Eq. (4) produces:

$$0 = \left(-\frac{w^2 f^2}{c^2} + \frac{P_g^2}{\hbar^2 c^2} - \frac{2wfP_g}{\hbar c^2}\right) r^{2g} + \left(\frac{2P_g P_{g-1}}{\hbar^2 c^2} - \frac{2wfP_{g-1}}{\hbar c^2}\right) r^{2g-1} + \left[\frac{P_{g-1}^2}{\hbar^2 c^2} + f^2(g^2 - k^2)\right] r^{2g-2}. \tag{6}$$

The problem here is overconstrained: we have three conditions in only two unknowns, $P_g$ and $P_{g-1}$. Despite the overconstraint, it is easily confirmed that the conditions can be satisfied by:

$$P_g = \left(1 \pm \sqrt{2}\right) f\hbar w, \text{ and } P_{g-1} = \sqrt{k^2 - g^2} = 0, \tag{7}$$

or:

$$P(r) = \left(1 \pm \sqrt{2}\right) f\hbar w r^{\pm k}. \tag{8}$$

So the potential in Eq. (1) goes as:

$$H_I = \left(1 \pm \sqrt{2}\right) f\hbar w r^{\pm k} \exp[i(k\vartheta - wt)]. \tag{9}$$

Thus of mathematical necessity, only the form of attractive potential defined by Eq. (9) can produce the exact solution (3). We can now apply conditions (7) and (8) to the rest of Eq. (4) and seek a complete solution. We will see shortly that the first and second brackets in Eq. (4) also place constraints on the solution, but first we impose the requirement that the wave function be square-normalizable: this requirement will allow us to evaluate $M(r)$, which appears in those brackets.

For that purpose, the quantity that takes the place of the probability density here is the charge density for the Klein-Gordon field[6]:

$$\rho = i(\varphi^{\dagger}\varphi_t - \varphi\varphi_t^{\dagger}), \tag{10}$$

where in the usual way the dagger and subscript represent the adjoint and time differentiation respectively. Using $\varphi$ from Eq. (3), we obtain $\rho \sim M^2(r)\, e^{2f\cdot r^{\pm k}cos(k\vartheta - wt)}$. If we attempt a power-law_solution, $M(r) = M_0 r^{\alpha}$, then the integrated density, $\iint \rho r dr d\vartheta \sim \iint r^{2\alpha+1} dr d\vartheta$, plainly contains an integrand that diverges at the origin for $\alpha < -½$ and at infinity for $\alpha > -½$. To prevent these divergences, we set $\alpha = -½$, so that:

$$M(r) = \frac{M_0}{\sqrt{r}}. \tag{11}$$

This prevents divergence due to the $M^2(r)$ term in the density, but notice that the remaining density, $\rho \sim ({M_0}^2/r)\, e^{2f\cdot r^{\pm k}cos(k\vartheta - wt)}$ still diverges as $r \to 0$ (if $k < 0$) or as $r \to \infty$ (if $k > 0$). Thus we require that $k = 0$, and so by Eq. (7), $g = 0$. This means that:

$$\varphi = \frac{M_0}{\sqrt{r}} \exp\{i(\pm½\vartheta - wt) + f\mathrm{e}^{-iwt}\}. \tag{12}$$

Additionally, since $k = 0$, our simple ansatz that the potential might rotate according to Eq. (1) is evidently overly general: the potential may oscillate in time, but not in space if the solution is to remain finite. The *solution*, Eq. (12), however is periodic azimuthally as well as in time, and – again similar to bouncing droplets – both potential and solution are synchronized in time: the essence of the dynamical *ansatz* that we began with.

Eq. (12) avoids divergences of the integrand, but not of the integral, which is the relevant quantity. To solve for the integral, we resume our analysis of the first and second brackets of Eq. (4), armed with $M(r)$ from Eq. (11).

Using Eq. (11), the first curly bracket in Eq. (4) becomes:

$$\frac{3}{4}r^{-5/2} - \frac{1}{2}r^{-5/2} - \left[\frac{w^2}{c^2} + \frac{\mu^2 c^2}{\hbar^2}\right] r^{-1/2} - \hat{\kappa}^2 r^{-5/2} = 0. \tag{13}$$

This must hold for all $r$, so collecting $r^{-5/2}$ terms, we obtain:

$$\hat{\kappa} = \pm½, \tag{14}$$

and the $r^{-1/2}$ term gives:

$$\hbar^2 w^2 + \mu^2 c^4 = 0. \tag{15}$$

## 3: Exact solution including feedback between potential and particle

The solution we have obtained so far includes the wave function (12), and a potential from Eq's (1) & (9):

$$H_I = n f \hbar w e^{-iwt}, \tag{16}$$

where $n = (1 \pm \sqrt{2})$. If we insert these results into the Klein-Gordon equation, we obtain a residual (i.e. a term on the right hand side of Eq. (2)) given by:

$$\frac{c^2\mu^2}{\hbar^2} + \frac{(1 - 2f^2 e^{2iwt})w^2}{c^2}. \tag{17}$$

The origin of this residual is straightforward: the first and third brackets in Eq. (4) have been satisfied, but it is easily confirmed that the second bracket is satisfied by $P_g = f\hbar w$ rather than by $P_g = (1 \pm \sqrt{2})f\hbar w$. This is to say that our luck seems to have run out: our solution provides a unique potential, Eq. (8), a unique form for the wave function, Eq. (12), and satisfies the over-constraint on Eq. (6), as well as 2 out of 3 brackets in Eq. (4), but there is a numerical discrepancy between the first and second of these brackets. More generally, for an arbitrary constant, $n$, in Eq. (16), the residual in the Klein-Gordon equation becomes:

$$\frac{c^2\mu^2}{\hbar^2} + \frac{1+e^{itw}f\,w^2\{3-3n-fe^{itw}[-1+n(2+n)]\}}{c^2}. \tag{18}$$

All of these results are strictly mathematical. We can, fortunately, avail ourselves of one further degree of freedom with a *physical* origin. The caricature of Fig. 1(b) depicts each particle periodically emitting (and then absorbing) a pilot wave. This is only a caricature, and as we mentioned in the introduction it neglects nonlinear interactions. Nevertheless, the pilot wave can do work on its neighbors, which as we have explained is why we chose a complex potential, Eq. (1). If we take the perspective that this work can only come out of the originating particle, then we might impose the physical deduction that the particle's rest mass, $\mu$, may vary periodically. We note that the notion that rest mass could oscillate is not new, and was in fact previously proposed by de Broglie[7].

This is the physical interpretation; mathematically we observe that the residual (18) would be zero if the mass, $\mu$, were to oscillate periodically (from whatever physical cause) according to:

$$\mu = \frac{\hbar w}{c^2}\sqrt{\{3(n-1) + fe^{itw}[n(n+2)-1]\}fe^{itw} - 1}. \tag{19}$$

Eq's (12), (16) & (19) then produce an exact solution to the Klein-Gordon equation in a periodic potential for arbitrary $M_0, w, n$ and $f$. Moreover, the integrand of $\iint \rho r dr d\vartheta$ is now constant, and diverges neither at small nor large $r$. As a consequence of being constant, however, the integral, $\int_0^R \rho r dr$, itself grows linearly with $R$, so Eq. (12) is still not square-normalizable. This is confirmed by direct calculation in the "monopole" inset to Fig. 2(a), where we plot the integral $\int_0^R \rho r dr$ obtained from Eq. (12) with $M_0 = f = 1$, $\vartheta = 0$, and $t = \pi/2w$. Since the integrand is constant, the numerically calculated integral predictably grows linearly with integration radius, $R$.

A remedy is available, however: a *single* wave function defined by Eq. (12) requires infinite energy, but *multiple* wave functions defined by this equation (i.e. multipoles) can be made square-normalizable. The simplest of these is a dipole:

$$\varphi_{dipole} = M_0 \exp(-iwt + fe^{-iwt})\left[\frac{\exp(\pm i\vartheta_+/2)}{\sqrt{r_+}} - \frac{\exp(\pm i\vartheta_-/2)}{\sqrt{r_-}}\right], \tag{20}$$

where the coordinates $(r_+, \vartheta_+)$ and $(r_-, \vartheta_-)$ define distances from the centers of two spatially separated wave functions. In Fig. 2(a), we directly calculate the integrated density, $\int_0^R \rho r dr$, of this "dipole" wave function exactly as we did for a monopole, and similarly we calculate the integrated density for a "tripole" consisting of an equilateral triangle of wave function centers, two with magnitude $M_0 = 1$, and a third with magnitude $M_0 = -2$. We plot the real parts of the corresponding wave functions in Fig. 2(b).

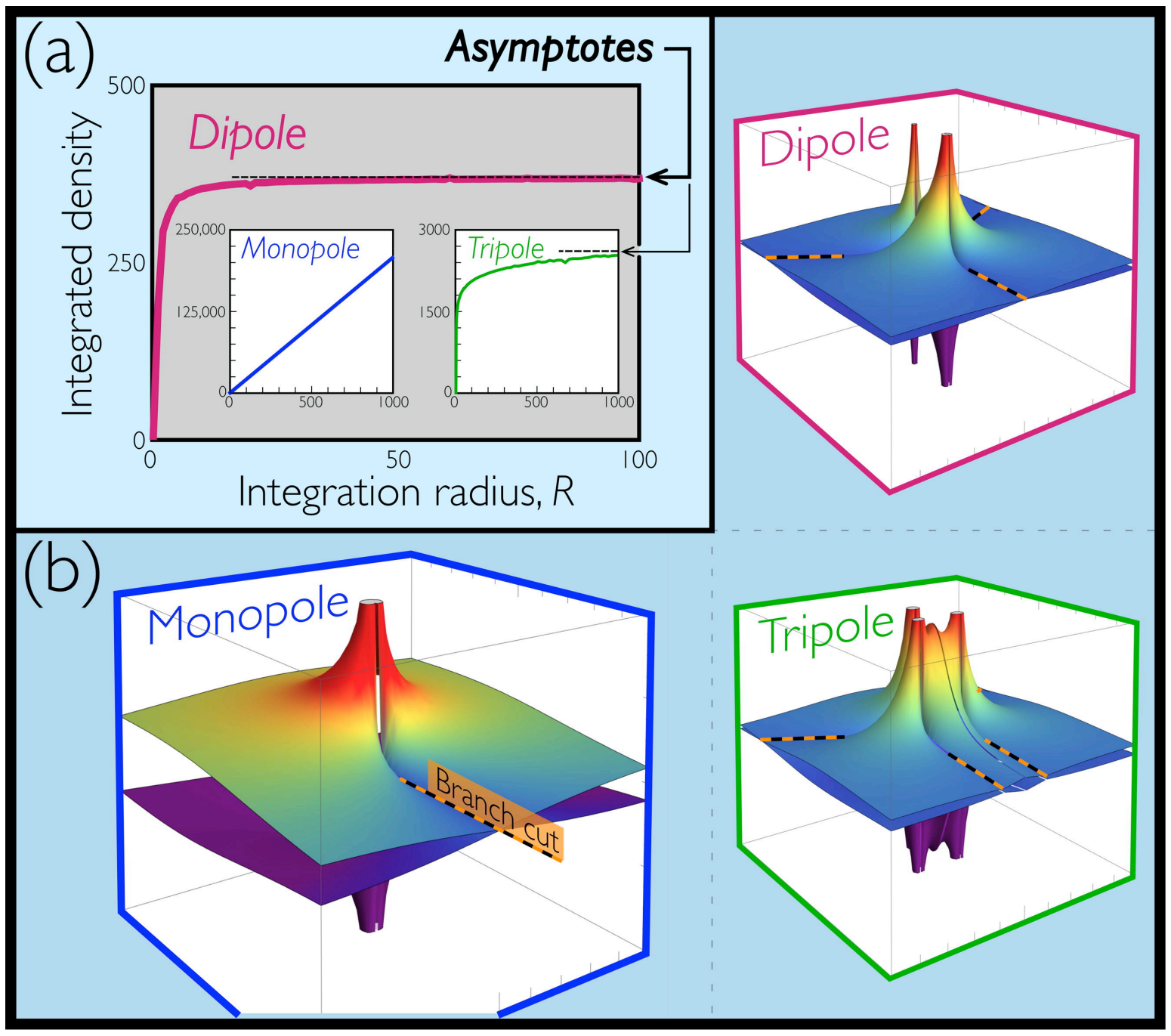

*Figure 2 – Wave function solutions at $M_0 = f = 1$ and time $t = \pi/2w$. (a) Integral, $\int_0^R \rho r dr$, using $\vartheta = 0$ as a function of radius, R, for monopole dipole, and tripole, where $\rho$ is the density (10). Notice that the monopole grows linearly without bound, while the dipole and tripole asymptote to finite values referred to in text. (b) Plots of $Re(\varphi)$ $vs. x$ & $y$. All plots are cut off at $Re(\varphi) = \pm 5$; the domain of the monopole is x,y ∈ $[-0.05, 0.05]$, while the dipole and tripole use x,y ∈ $[-5, 5]$. Branch cuts discussed in text are highlighted.*

The calculations showing that $\int_0^R \rho_{dipole} r dr$ and $\int_0^R \rho_{tripole} r dr$ are finite were performed in Fig. 2(a) for $\vartheta = 0$, but any other choice of $\vartheta$ would merely include a multiplicative factor $e^{\pm i\vartheta}$. Consequently, all choices of $\vartheta$ result in finite densities, and so an integral over $\vartheta$ also produces finite amplitudes. If we call that amplitude $A_{multipole}$, then we can normalize any multipole by defining its wave function magnitude to be:

$$M_0 = 1/\sqrt{A_{multipole}}. \tag{21}$$

This is unusual in that the magnitude of a dipole may be different from the magnitude of a tripole; nevertheless it allows us to define normalized wave functions, i.e. $\langle\varphi|\varphi\rangle = 1$, in the usual way. Additionally, we note for completeness that probability is dimensionless, so the square of a 2D wave function must have dimension $L^{-2}$, and the function itself must have dimension $L^{-1}$. Our solution, Eq. (12), provides a wave function with dimension $M_o L^{-1/2}$, so $M_o$ cannot be dimensionless. Dimensional constants are not remarkable, but the consequence in this case is that the amplitude of the wavefunction, $M_o$, necessarily carries a lengthscale. We cannot speak to the physical origin of this lengthscale, but since the bound states rely on a potential, Eq. (16), and mass, Eq. (19), both of which oscillate with frequency $w$, a natural lengthscale for the Klein-Gordon problem is $c/w$.

## 4: Discussion:

We now have in hand a wave function (12) that solves the Klein-Gordon equation in 2D in the presence of a periodic potential (16) and oscillating mass (19). In order to produce this solution, we have seen that several necessary mathematical constraints arise: we enumerate some physical consequences of these constraints here.

First, we can calculate the angular momentum of our solution:

$$L_z = \left\langle \varphi \middle| -i\hbar \frac{\partial}{\partial \vartheta} \middle| \varphi \right\rangle \equiv \pm \frac{\hbar}{2}, \tag{22}$$

where again we normalize $\varphi$ according to Eq. (21). This is unexpected, for it reveals that the angular momentum of a periodic dynamical state has half-integer angular momentum – despite the fact that the Klein-Gordon equation applies to zero spin. Nevertheless, this is the finding, which arises from the simple $e^{\pm i\vartheta/2}$ term in Eq. (12) – and whose origin is in the requirement of a finite density. In a sense, this is a mathematical take on the meaning of "intrinsic" angular momentum: nothing constrains solutions to have any particular value of physical angular momentum; finite solutions simply don't exist unless $L_z = \pm\hbar/2$. *Thus integer spin particles can produce half-integer angular momentum states.*

An incidental consequence of the same, $e^{\pm i\vartheta/2}$, dependence is that solutions are periodic in $4\pi$, and so branch cuts necessarily arise: these are identified in Fig. 2(b). This dependence of course implies that after one circuit of $2\pi$, the wave function appears to be out of phase with itself: a seeming failure of single-valuedness. To make sense of this, we therefore must consider the wave function to live on a Riemann surface of periodicity $4\pi$. This finding can be viewed in the light of the double-coverage of SO(3) by SU(2), but we reiterate that we have not produced our result from a new physical or quantum-mechanical hypothesis. The physics of double-coverage in quantum mechanics is old news and is not our concern here: our result arises as a strictly mathematical outcome. *Thus a doubly-periodic solution arises naturally from the Klein-Gordon equation with the singly-periodic periodic potential Eq. (16).*

Second, we found that a single solution cannot exist alone, but must co-exist with one or more others of the same form. We once more reiterate that our results are only mathematical; nevertheless it would be difficult not to recognize a physical parallel with quark confinement. Our solutions cannot exist in isolation because such a solution is not finite (recall Fig. 2(a), monopole inset). This differs in *physical* significance from traditional descriptions of confinement in that it is not necessary to make any new assumptions – e.g. bag forces that grow with separation, unusual perturbative expansions, gluon chains, etc. – it is simply the case that a single solution (12) by itself diverges, but composing solutions as in Eq. (20) can produce a finite outcome. *Thus confinement of solutions appears for free as a necessary consequence of our construction as well.*

Third, note that the solution shown relies on a complex term in Eq. (1). Complex quantum problems have previously been described[8,9], and it is easily shown that the physical significance is that such a system has oscillatory energy – i.e. $\left\langle \varphi \middle| -i\hbar \frac{\partial}{\partial t} \middle| \varphi \right\rangle$, is not constant. Indeed, we have mentioned that our potential represents an oscillatory input in energy – analogous to the situation of the bouncing droplet, but applied here to quantum waves; here also we have mentioned that rest mass oscillations in solutions to the Klein-Gordon equation have previously been described[10]. *Thus the exact solution that we have described requires periodic transfer of energy in both directions between particles and their pilot waves.*

Finally, we find that time-dependent oscillations in potential, Eq. (16), and mass, Eq. (19), produce solutions, Eq. (12), that circulate at phase speed $w/2$, and only one particular form of mass fluctuation, Eq. (19), leads to an exact solution. Interactions between the solution, the potential and the particle mass involve unexpectedly complicated dynamics, shown in the mass fluctuations whose real part we plot in Fig. 3 for several choices of $n$, all using $w = f = \hbar = c = 1$.

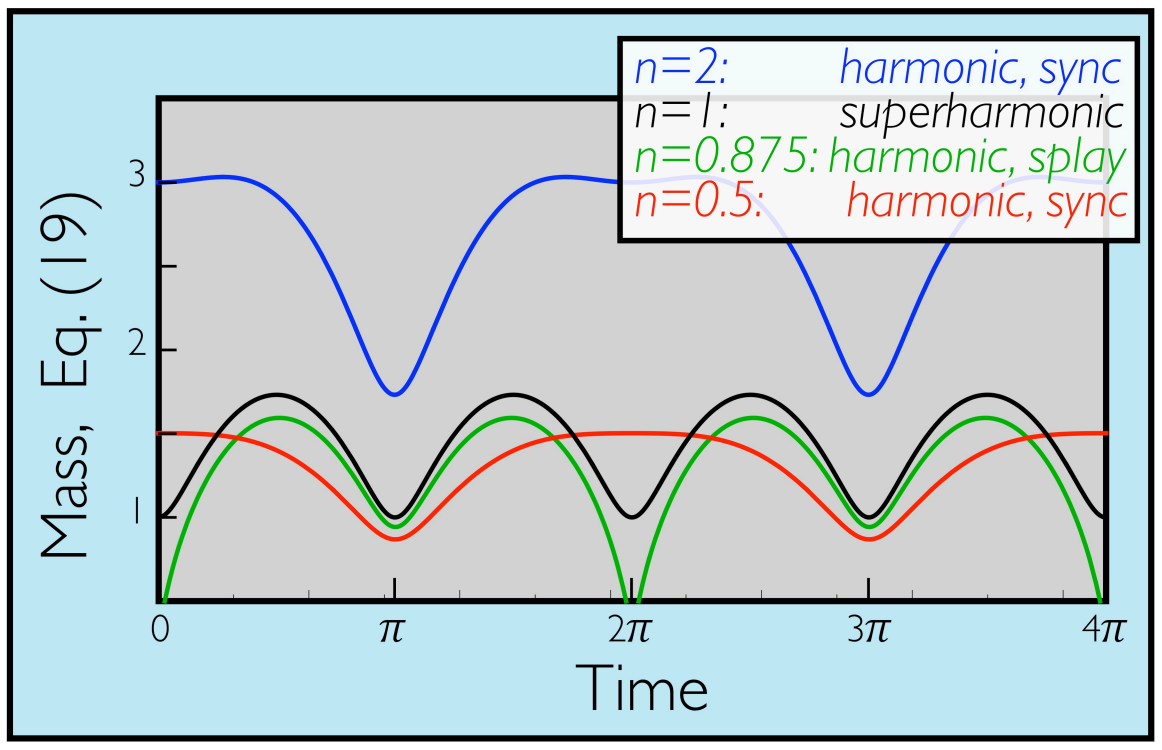

*Figure 3 – Periodic mass fluctuations from the real part of Eq. (19) associated with various choices of the strength, n, of the potential defined by Eq. (16). Notice that some states are superharmonic (twice as frequent as the potential's periodic fluctuations), some are dominantly in phase ("sync"), meaning mass decreases whenever the potential increases), while others in a narrow range, 0.73 ‹ n ‹ 1, are dominantly out of phase between mass and potential ("splay").*

At its simplest, the mean mass increases with the strength, $n$, of the potential. Additionally, for $n = 1$, each solution loses mass twice as fast as $wt$ (labeled "superharmonic" in the figure). By comparison, for $n$ far from 1 (i.e. for either very weak or very strong potentials), remnants of superharmonic oscillations persist, but predominantly each solution loses and gains mass once per $wt$ ("harmonic"), and the mass decreases as the real part of the potential increases ("sync"). This is amenable to the physically logical interpretation that emission of a pilot wave that produces the potential costs a particle mass. Finally, for $n$ between about 0.73 and 1, the solution again shows superharmonic oscillations, but is predominantly harmonic with mass is $\pi$ out of phase with the potential (termed a "splay" state in the literature[11]). Purely speculatively, this could be interpreted as being associated with a field whose influence takes time $t = \pi/w$ to travel from emission to absorption, resulting in mass growth that instantaneously coincides with potential growth. It is also worth noting that the square in the Klein-Gordon equation (Eq. (2)) implies that the driving function itself contains both harmonic and twice-harmonic terms.

It remains to be seen if any of these mathematically predicted states have physical consequences, but dynamical features of solutions are evidently nontrivial. Apparently, periodic driving may produce multiple distinct dynamical states: some superharmonic, some dominantly harmonic, some symmetric with the pilot wave, others antisymmetric. These dynamical features can only be richer in a fully nonlinear – and possibly 3D – model.

As a final caveat, recall that we began by imagining that a central oscillatory potential (Eq. (1)), could generate localized bound states. We found that exact localized solutions do in fact exist, but the potential must be spatially uniform, and can only oscillate in time (Eq. (16)). The emergence of localized bound states from a uniform forcing potential is unexpected, but it is a result of mathematical analysis, and is unalterable.

In support of the possibility of physical analogs of this mathematical solution, we note that the droplet problem does exhibit localized states that arise from spatially uniform oscillatory forcing. Indeed, both droplets and the Klein-Gordon equation produce localized states through interactions between oscillating masses and spatiotemporal waves, not through a central potential. Beyond this, possibly only serendipitous, correspondence, mathematical analysis cannot resolve physical questions. For example, our solution relies on an oscillatory potential, but one can only speculate whether such oscillations are actually present in nature, or whether oscillating masses above a critical density might synchronize to spontaneously generate such a potential. Our goal here has not been to declare that Klein-Gordon bound states actually exist in nature, it has only been to demonstrate that such solutions are mathematically possible.

## Conclusion:

Motivated by intriguing parallels between bouncing droplet dynamics and quantum phenomena, we have introduced a periodic potential into a standard quantum wave equation. We found that an exact solution can be produced by straightforward derivation combined with the *ad hoc* manufacture of a term that cancels a residual and that can be interpreted as mimicking feedback between solutions and pilot waves on which they depend. We reiterate that the potential and mass oscillations required to generate an exact solution are only *ad hoc*, and do not represent the full nonlinear behavior of this feedback. So these findings fall short of providing a nonlinear analysis to fully describe quantum states bound by a periodic potential.

Nevertheless, it is worth reflecting that in 1963 Ernest Abers and others[12,13] advocated in related contexts that masses and interactions of particles "need not be regarded as a freak accident" but could result from a "self-consistent calculation." If the results of calculations here are to be believed, several empirical observations may all be derivable from straightforward mathematics. These include the possibility that spin ½ states may be generated by integer spin particles and that other properties such as confinement may have mathematical origins. Additionally, new behaviors of bound states have been derived, including superharmonic states and states with masses either synchronized or anti-synchronized with oscillations in potential. All of these properties arise from basic algebra constrained by square-normalizability, and not from any new physical hypothesis. The existence of exact solutions for our *ansatz*, Eq. (1), suggests that other exact solutions – potentially with their own physical parallels – could also be found.

Acknowledgments: I acknowledge support from the NSF DMR, award #1404792, and NSF CBET, award #1804286, and I thank Daniel Harris and John Bush for their droplet photograph and helpful comments; I also thank an anonymous reviewer for cogent remarks critical to the solution presented.